\documentclass[amsmath,amsfonts,preprint,prb,aps,tightenlines]{revtex4}

\usepackage{epsfig}

\newcommand{\tens}[1]{\mbox{\sffamily\bfseries{#1}}}

\def\BEQ {\begin{equation}} 
\def\EEQ {\end{equation}}
\def\BEQAR {\begin{eqnarray}}
\def\EEQAR {\end{eqnarray}}

\def\En#1{\label{Eq.#1}}
\def\E#1{(\ref{Eq.#1})}
\def\Rn#1{#1}
\def\Ron#1{\onlinecite{#1}}
\def\R#1{\cite{#1}}
\def\Fn#1{\label{Fig.#1}}
\def\F#1{\ref{Fig.#1}}

\def\Sn#1{\label{Sec.#1}}

\def\hor #1{{\hskip #1 mm}}
\def\ver #1{{\vskip #1 mm}}

\def\bfxi {{\boldsymbol{\xi}}}

\def\bfb {{\bf b}}

\def\bfk {{\bf k}}

\def\bfv {{\bf v}}

\def\bfF {{\bf F}}

\def\bfV{{\bf V}}

\def\half {{\textstyle \frac{1}{2}}}

\def\div #1{\nabla\cdot #1}

\def\vvvph {\vphantom{\Bigg(}}

\def\ds{\displaystyle}

\begin{document}

\title{Transonic instabilities in accretion disks}

\author{J.P. Goedbloed and R. Keppens}
\affiliation{FOM-Institute for Plasma Physics `Rijnhuizen', Nieuwegein \\
\& {Astronomical Institute Utrecht} \\
$\quad$goedbloed@rijnh.nl}

\date{\today}

\begin{abstract}

In two previous publications~\R{KSG02, GBHK04a}, we have demonstrated that stationary rotation of magnetized plasma about a compact central object permits an enormous number of different MHD instabilities, with the well-known magneto-rotational instability~\R{Vel59, Chan60, BH91} as just one of them. We here concentrate on the new instabilities found that are driven by transonic transitions of the poloidal flow. A particularly promising class of instabilities, from the point of view of MHD turbulence in accretion disks, is the class of {\em trans-slow Alfv\'en continuum modes}, that occur when the poloidal flow exceeds a critical value of the slow magnetosonic speed. When this happens, virtually every magnetic/flow surface of the disk becomes unstable with respect to highly localized modes of the continuous spectrum. The mode structures rotate, in turn, about the rotating disk. These structures lock and become explosively unstable when the mass of the central object is increased beyond a certain critical value. Their growth rates then become huge, of the order of the Alfv\'en transit time. These instabilities appear to have all requisite properties to facilitate accretion flows across magnetic surfaces and jet formation. 

\end{abstract}


\maketitle

\section{Introduction}{\Sn{1}}

In Fig.~\F{1}, a {\em Magnetized Accretion-Ejection Structure} is shown which illustrates the problem we wish to address in this paper, viz.: How does an accretion flow about a compact object first crosses the magnetic configuration and then turns the corner with respect to the accretion disk to produce jets? In ideal MHD, plasma and magnetic field stay together (frozen in field), so that a sizeable resistivity is needed for the flow to detach from the magnetic field. This involves anomalous dissipation. Hence, the basic problem is to find relevant {\em local instabilities} producing the necessary MHD turbulence.
\begin{figure}[ht]
\begin{center}
\hor{10}\includegraphics[height=8cm]{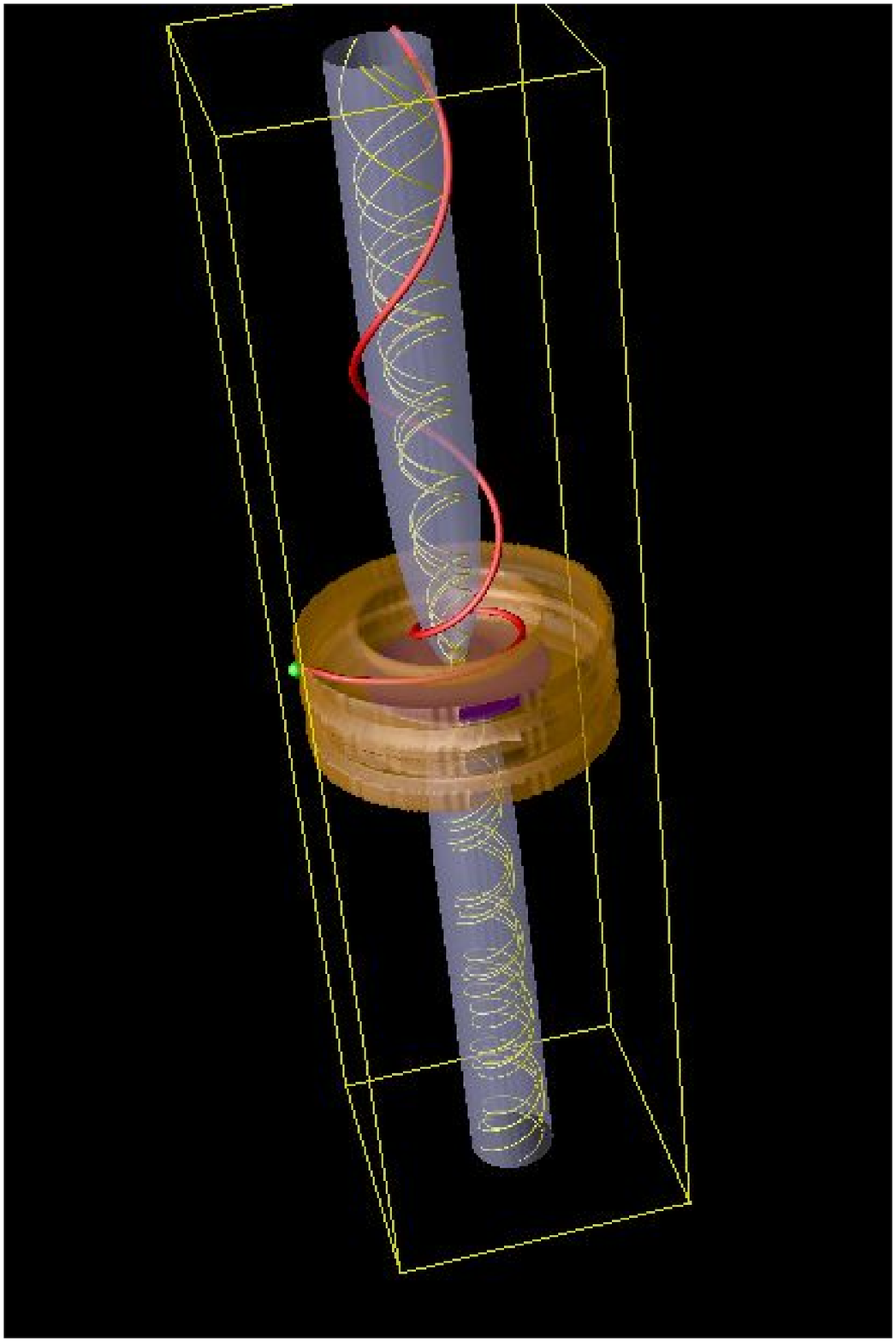}\hor{2}{\raisebox{-6mm}{\includegraphics[width=9.5cm]{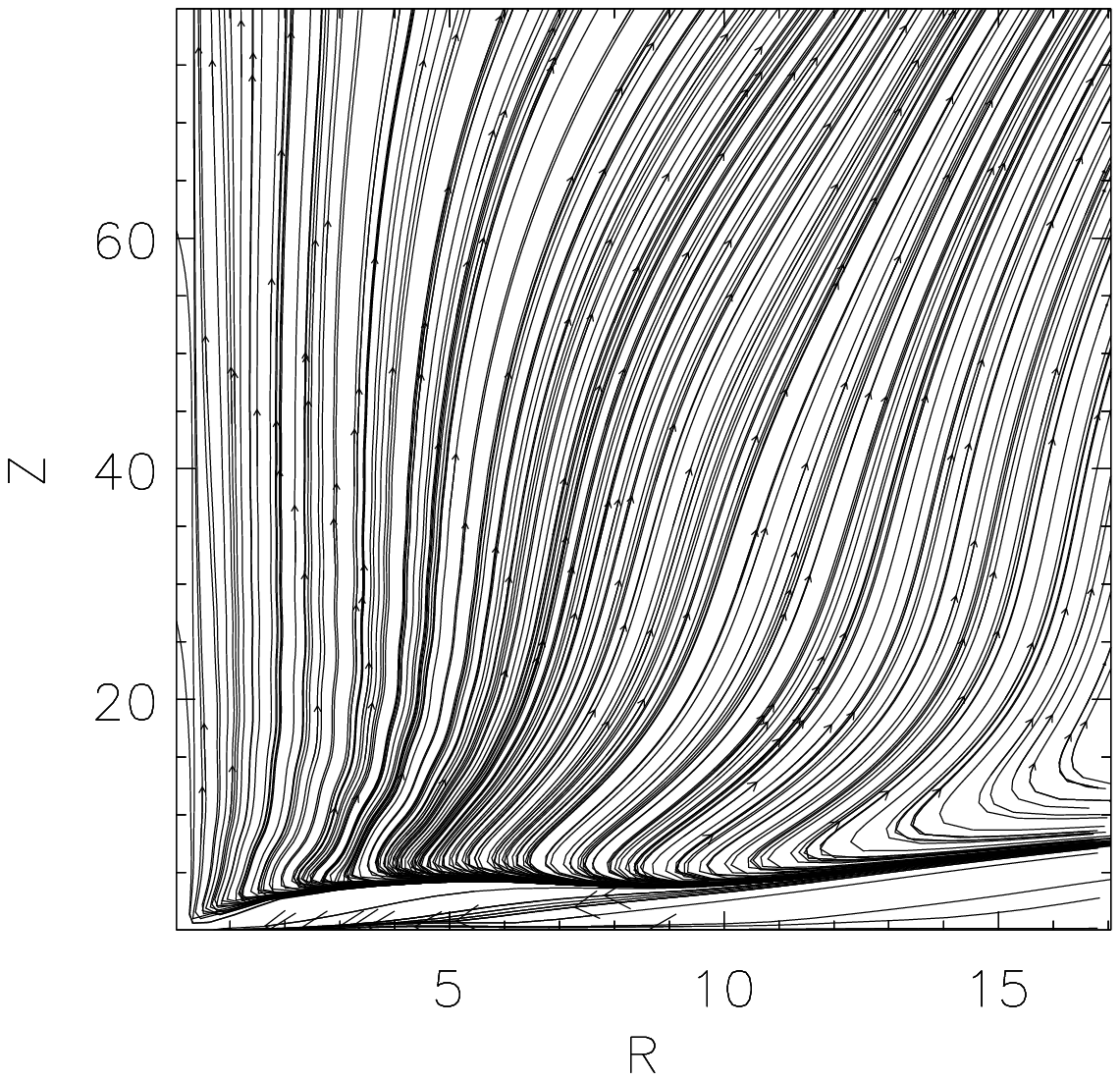}}}
\end{center}
\vspace{-4mm}
\caption{(a)~Stationary end state from simulation with VAC~\R{Toth96} (Versatile Advection Code): disk density surfaces (brown), jet magnetic surface (grey), helical field lines (yellow), accretion-ejection particle trajectory (red);
(b)~Accretion flow detaching from the disk. [$\,$From computations by Casse and Keppens~\R{CK04, CK02}$\,$].\Fn{1}}
\end{figure}

Our model is shown in Fig.~\F{2}: An axisymmetric configuration of nested magnetic$\,/\,$flow surfaces with magnetic field indicated by the vectorial Alfv\'en speed $\bfb$ and velocity $\bfv$, having both toroidal and poloidal components, surrounds a compact object of mass $M_{\textstyle*}$ in the origin. Note that the usual tokamak configuration is obtained for $M_{\textstyle*} = 0$, whereas accretion disk geometries may have flat (thin disk) as well as round (thick disk) poloidal cross-sections.~\R{FKR92} This model considers laboratory and astrophysical toroidal plasmas on an equal footing by exploiting the {\em scale independence}~\R{GP04} of the MHD equations.  
\begin{figure}[ht]
\begin{center}
\includegraphics[width=6cm]{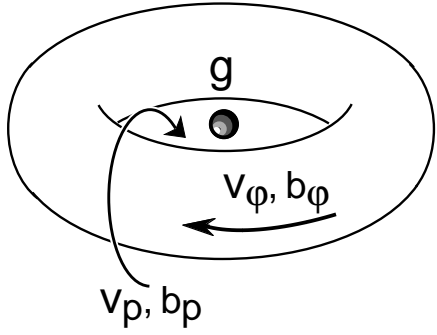}
\end{center}
\vspace{-4mm}
\caption{Transonically rotating magnetized disk about compact object. [$\,$From Ref.~\Ron{GBHK04a}$\,$].\Fn{2}}
\end{figure}

In order to obtain a stationary  equilibrium situation, we assume that the accretion flow speed is much smaller  than both rotation speeds of the disk. We then need to determine the {\em stationary equilibrium flows} (Sec.~2) and, next, the {\em local instabilities driven by the transonic flow} (Sec.3). We analyze this problem from two angles:\\
(a)~Asymptotic analysis for {\em small inverse aspect ratio} ($\epsilon \ll 1$);\\
(b)~Large-scale {\em exact numerical} computations of the equilibria and instabilities.\\
These will be discussed in reverse order since the numerical results suggest the relevant approximations that may be made. 

A difficulty encountered is that {\em transonic transitions upset the standard equilibrium--stability split}. We will discuss this in Sec.~2 under the heading of transonic enigma.

The {\em gravitational parameter} exploited here is defined as follows: 
\BEQ \Gamma(\psi) \equiv \frac{\rho G M_{\textstyle*}}{R_0 M^2 
B^2} \quad\left[\,\approx \frac{G M_{\textstyle*}}{R v_\varphi^2} \;\; \hbox{{\em for parallel flow}} \right] \,. \En{1}\EEQ
This parameter measures the deviation from Keplerian flow (for which $\Gamma = 1$).

\section{Transonic equilibrium flows}{\Sn{2}}

We exploit a {\bf variational principle for the stationary axisymmetric equilibria} determining the poloidal flux $\,\psi\,$ and the poloidal Alfv\'en Mach number squared $\,M^2 \ \big(\equiv \rho v_p^2 / B_p^2\big).$ This involves five arbitrary scaled flux functions $\Lambda_i(\psi)\,$:
\BEQAR
\chi'(\psi)&\equiv& \rho v_p / B_p \,, \hor{29}\hbox{\em (derivative poloidal stream function)} \En{2}\\[1mm]
H(\psi) &\equiv& \half v_p^2 \,\frac{B^2}{B_p^2} + \frac{\gamma}{\gamma-1} 
\,\frac{p}{\rho} - \half R^2 \Omega^2 - 
\frac{G M_{\textstyle*}}{\sqrt{R^2 +Z^2}} \,, \hor{7}\hbox{\em (Bernoulli function)} \En{3} \\[0mm]
S(\psi) \hor{1}&\equiv& \rho^{-\gamma} p \,, \hor{33}\hbox{\em(entropy)} \En{4}\\[3mm]
\Omega (\psi) \hor{1}&\equiv& {R}^{-1} \big[\, v_\varphi - (\chi' / \rho) 
\,B_\varphi \big] \,, \hor{4}\hbox{\em($-$derivative electric potential)} \En{5} \\[4mm]
K(\psi) &\equiv& R \,\,\big[\, v_\varphi - (1 / \chi') B_\varphi \big] \,,\hor{8}\hbox{\em(pol.~vorticity \& current density stream function)}\En{6}
\EEQAR
which have to be fixed by whatever observational evidence is available. Note that the flux function~\E{6} has been renamed $K$ (instead of the usual $L$) to eliminate the confusion derived from the frequent occurrence of the misnomer `specific angular momentum' in  the astrophysics literature (see references in Ref.~\Ron{GBHK04a}). This is important since one of the essential problems in accretion disk dynamics is precisely the transport of angular momentum.

The stationary states are then determined by minimization of the following Lagrangian:
\BEQ \delta \int {\cal L} \,dV = 0 \,, \quad {\cal L} \equiv \frac{\displaystyle 1}{\displaystyle 2R^2} (1 - M^2) |\nabla \psi|^2 - \frac{\displaystyle\Pi_1}{\displaystyle M^2} - \frac{\displaystyle\Pi_2} {\displaystyle\gamma M^{2\gamma}} 
+ \frac{\displaystyle\Pi_3}{\displaystyle 1 - M^2} \,,\En{7}\EEQ
where $\Pi_j(\Lambda_i(\psi); R, Z)\,$ are simple algebraic expressions. The {\em Euler equations} provide the solutions $\psi(R,Z)$ and $M^2(R,Z)\,$ of the core variables. [$\,$A generalization of this variational principle to two-fluid plasmas is given in Ref.~\Ron{GBHK04c}.$\,$]

{\bf The transonic enigma} mentioned above is due to the fact that the flows suddenly change character from elliptic to hyperbolic at the transonic transitions. As a result, {\em standard (tokamak) equilibrium solvers diverge in the hyperbolic regimes!} We circumvent this problem by calculating in elliptic regimes beyond the first hyperbolic one. Obviously, the payoff is that we cannot approach the transonic transitions but have to infer what has happened there from the changes in the dynamics found in the `transonic' elliptic regimes. 

The pleasing side of the transonic enigma is that the time-dependence of the linear waves and the spatial dependence of the nonlinear stationary states are intimately related. This is seen by comparing the wave spectra, which cluster at the {\em slow, Alfv\'en, and fast continuum frequencies} $\{\omega_S^2\}$, $\{\omega_A^2\}$, $\omega_F^2 \equiv \infty$ for highly localized modes (Fig.~\F{3}),
\begin{figure}[ht]
\begin{center}
\includegraphics[width=12cm]{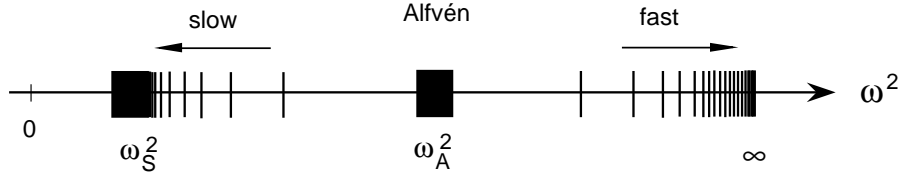}
\end{center}
\vspace{-4mm}
\caption{Cluster spectra of the waves.\Fn{3}}
\end{figure}
\noindent
and the corresponding {\em slow, Alfv\'en,  and fast hyperbolic flow regimes} delimited by critical values of the square of the poloidal Alfv\'en Mach number (Fig.~\F{4}):
\begin{figure}[ht]
\begin{center}
\includegraphics[width=12cm]{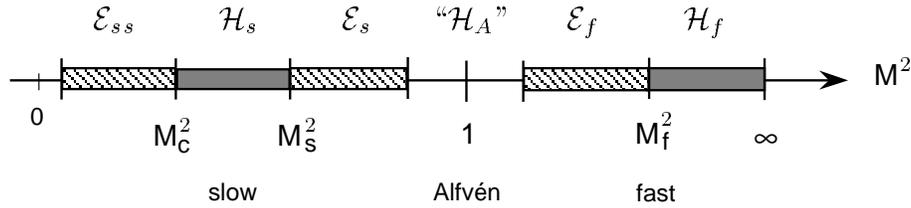}
\end{center}
\vspace{-4mm}
\caption{Flow regimes of the stationary states.\Fn{4}}
\end{figure}

By means of the transonic equilibrium solver FINESSE (described in Ref.~\Ron{BBGHK02}) typical {\bf equilibria in the first trans-slow elliptic regime} have been computed for tokamak and accretion disk (Figs.~\F{5}, and \F{6}) for a representative choice of the flux function parameters:
\begin{figure}[ht]
\begin{center}
\includegraphics[width=12cm]{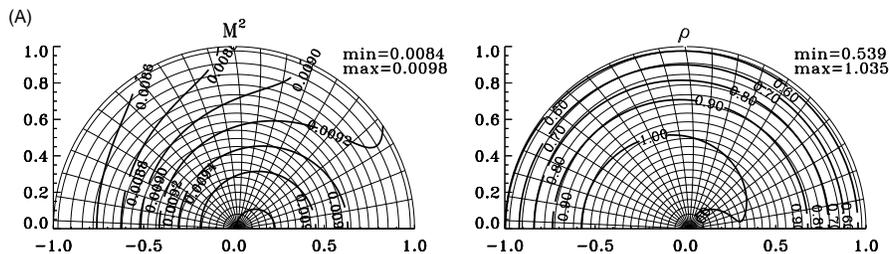}
\end{center}
\vspace{-4mm}
\caption{Equilibrium for tokamak ($\Gamma = 0$). \Fn{5}}
\end{figure}
\begin{figure}[ht]
\begin{center}
\includegraphics[width=12cm]{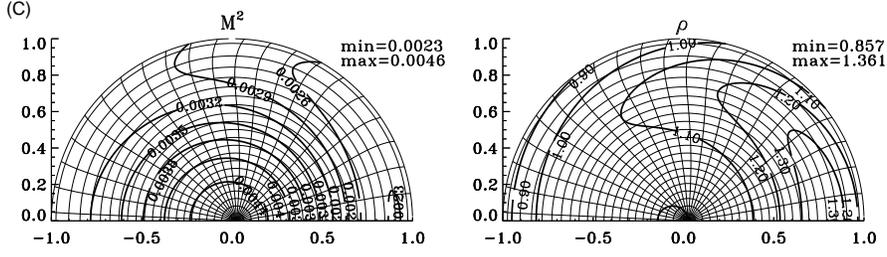}
\end{center}
\vspace{-4mm}
\caption{Equilibrium for accretion disk ($\Gamma = 2$). \Fn{6}}
\end{figure}

Note that the accretion disk equilibrium, in contrast to the tokamak, has the density peaking on the outside to produce overall equilibrium on the flux$\,/\,$flow surfaces with respect to the gravitational pull of the compact object in the center. For the sake of the spectral calculations (Sec.~3), the two equilibria have been chosen such that the safety factor $q(\psi)$ is a monotonically increasing function. 

\section{Local transonic instabilities}{\Sn{3}}

With equilibrium flows, the overall {\bf spectral structure} of MHD waves and instabilities is determined by the split in forward and backward waves so that the local waves cluster at the {\em Doppler shifted continuous spectra:} 
\BEQ\Omega_S^{\pm}= \pm \omega_S + \bfk \cdot \bfv\,,\qquad\Omega_A^{\pm} = 
\pm \omega_A + \bfk \cdot \bfv\,, \qquad\Omega_F^{\pm} = \pm \infty\,.\En{8}\EEQ 
These are embedded in a monotonic spectral structure for 1D equilibria, as shown in Fig.~\F{7}. A long-standing puzzle about the nature of the singular frequencies $\Omega_0$ ($\equiv \bfk \cdot \bfv$) has been clarified in Ref.~\Ron{GBHK04b}: In the Eulerian description, these frequencies give rise to the Eulerian entropy continua ($\Omega_E$), not perturbing the pressure, velocity, and magnetic field, and, hence, absent in the Lagrangian description.
\begin{figure}[ht]
\begin{center}
\includegraphics[width=12cm]{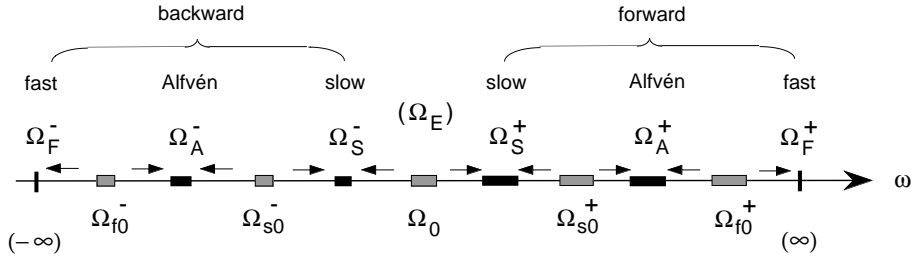}
\end{center}
\vspace{-4mm}
\caption{Schematic spectral structure for 1D stationary equilibria.~\Fn{7}}
\end{figure}

To solve for the local transonic instabilities, we exploit the {\em Frieman-Rotenberg equation,}
\BEQ \bfF_{\rm static}(\bfxi) + \div{\big[\rho (\bfv \cdot \nabla \bfv) \bfxi \big]} + \rho (\omega + i \bfv \cdot \nabla)^2 \bfxi = 0 \,.\En{9}\EEQ
The linear operator is no longer self-adjoint, so that {\em overstable modes occur}, in particular in 2D axisymmetric equilibria through {\em coupling of the poloidal modes $e^{im\vartheta}$}.

In order to solve for the {\bf transonic continuum modes} we exploit localization on separate magnetic$\,$/$\,$flow surfaces,
\BEQ\xi(\psi, \vartheta, \varphi) \approx \delta(\psi-\psi_0) \,\hat{\xi}(\vartheta) \,e^{i n \varphi} \,.\En{10}\EEQ
This gives rise to an eigenvalue problem for each surface, 
\BEQ \hbox{\vvvph$ \hat{\tens{A}} \cdot \hat{\bfV} = \hat{\tens{B}}\cdot \hat{\bfV}$} \,,\qquad \hat{\bfV} \equiv 
(\hat{\xi_\perp}\,, \hat{\xi_\parallel})^{\rm T} \,, \En{11}\EEQ
where  the matrices are defined by
\BEQAR 
\hor{-7}&&\hat{\tens{A}} \equiv \left( \begin{array}{cc}
\hor{-2}\ds\vvvph {\cal F} \frac{R^2B_p^2}{B^2}{\cal F} - \big(M^2 - M_c^2 \big) 
\frac{B^2}{\rho^2} \Bigg[ \partial \Big(\frac{\rho R B_\varphi}{B^2}\Big) \Bigg]^2 &\hor{-2}\ds - i (M^2 - M_c^2) \frac{B^2}{\rho^2} \Bigg[ \partial \Big(\frac{\rho R 
B_\varphi}{B^2}\Big) \Bigg] {\cal F} \rho \\[10mm]
\hor{-10}\ds i \rho {\cal F} (M^2 - M_c^2) \frac{B^2}{\rho^2} \Bigg[ \partial 
\Big(\frac{\rho R B_\varphi}{B^2}\Big) \Bigg] &
\hor{-13.5}\ds {\cal F} M_c^2 B^2 {\cal F} + \rho \Bigg[\partial \Big(\big(M^2 - M_c^2 \big)\frac{B^2}{\rho^2} \partial\rho \Big)\Bigg]
\end{array} \!\!\!\right)\!\!, \\\En{12}
\hor{-7}&&\hat{\tens{B}} \equiv \left( \begin{array}{cc}
\hor{-4}\ds \big(\sqrt{\rho} \,\widetilde \omega - {\cal F} M \big) 
\frac{R^2B_p^2}{B^2} \big(\sqrt{\rho}\,\widetilde \omega - M {\cal F} \big) &
\ds \hor{-20} \quad - i \alpha \sqrt{\rho} \,\widetilde \omega \\[10mm]
\ds \hor{-20} i \alpha \sqrt{\rho} \,\widetilde \omega &
\ds \hor{-10}\big(\sqrt{\rho}\,\widetilde \omega - {\cal F} M \big)B^2 
\big(\sqrt{\rho}\,\widetilde \omega - M {\cal F} \big)
\end{array} \!\right)\!\!, \En{13}\EEQAR
and the Doppler shifted frequency $\,\widetilde{\omega}\equiv \omega - n \Omega\,$ in a frame rotating with $\,\Omega \, (\,\ne v_\varphi/R\ !)\,$.

The overall result of the analysis of this eigenvalue problem~\E{11} is that the continuum modes are {\em always unstable in the trans-slow ($\,M^2 > M_c^2\,$) flow regime!} This is due to the poloidal derivatives indicated by the terms $[\partial \,(\ldots)]$ multiplied with factors $(M^2 - M_c^2)$. The instability mechanism involves coupling of the Alfv\'en and slow continuum modes.

Fig.~\F{8} shows the full complex spectrum of trans-slow Alfv\'en continuum `eigenvalues' for a {\bf Tokamak ($\Gamma =0$)} equilibrium with the radial coordinate $s \equiv {\psi}^{1/2}$ as a parameter. The colors indicate on which magnetic$\,/\,$flow surface the modes are localized. The modes rotate clockwise for Re$\,\bar{\omega} > 0$ and anti-clockwise for Re$\,\bar{\omega} < 0$, and have growth rates of the order of a few percent of the inverse Alfv\'en transit time. 
\begin{figure}[ht]
\begin{center}
\includegraphics[width=12cm]{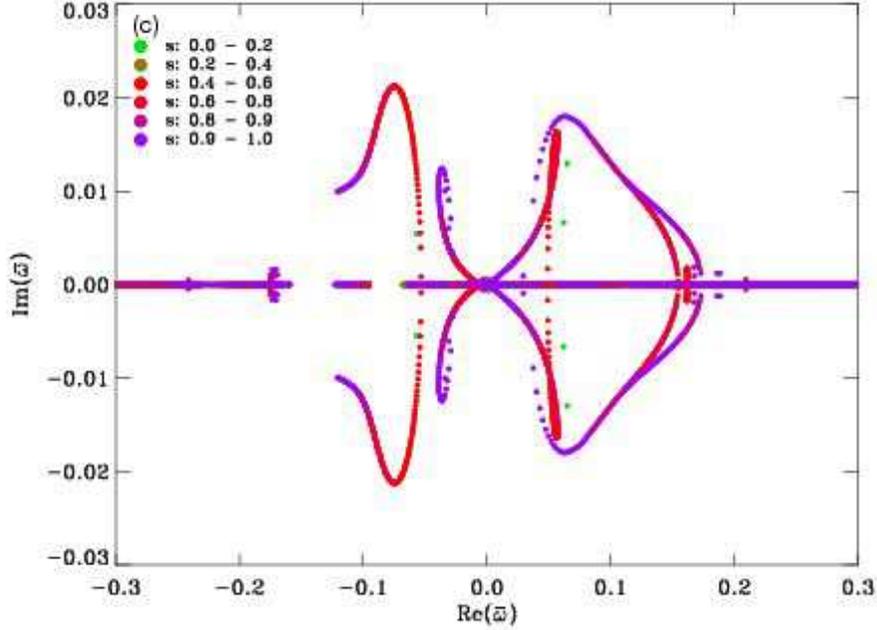}
\end{center}
\vspace{-4mm}
\caption{Complex eigenvalues of  $n=1$ transonic instabilities for tokamak ($\Gamma = 0$).~\Fn{8}}
\end{figure}

The counterpart for an {\bf accretion disk ($\Gamma =2$)} equilibrium is shown in Fig.~\F{9}. Now the rotating continuum modes lock to give purely exponential growing modes (Re$\,\bar{\omega} = 0$) over a sizeable range of magnetic$\,/\,$flow surfaces. Their growth rates are huge, in the order of ten to twenty percent of the inverse Alfv\'en transit time! Consequently, these modes have enough time to saturate during a finite number of revolutions of the plasma.
\begin{figure}[ht]
\begin{center}
\includegraphics[width=12cm]{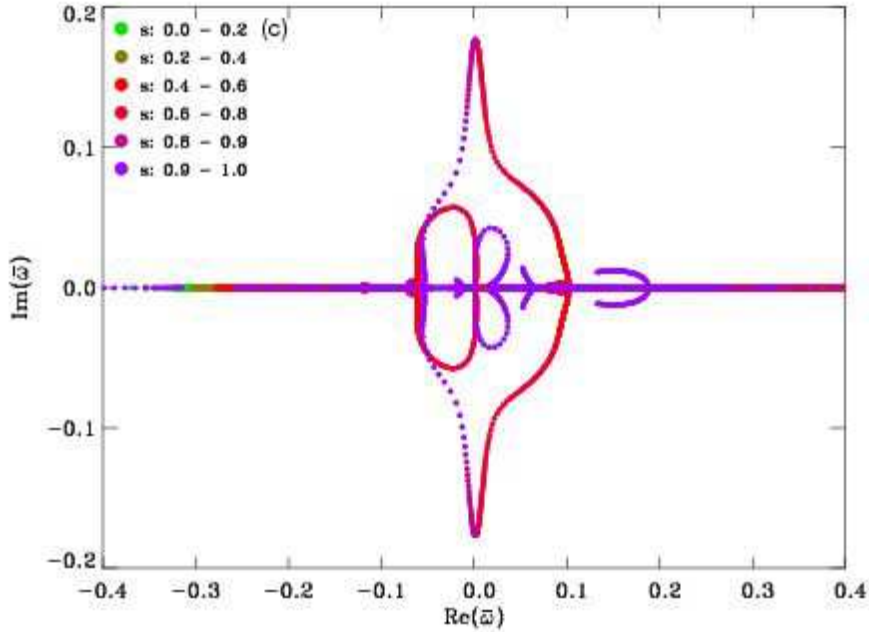}
\end{center}
\vspace{-4mm}
\caption{Complex eigenvalues of transonic  $n=1$ instabilities for accretion disk ($\Gamma =2$).~\Fn{9}}
\end{figure}

{\bf Analysis of the dispersion equation} by small inverse aspect ratio expansion ($\epsilon \ll 1$) is suggested by the numerical results, which exhibit dominant coupling of the six Alfv\'en and slow continuum modes $S_{m-1}^\pm$, $A_m^\pm$, $S_{m+1}^\pm$ around the $\epsilon = 0$ degeneracies at the rational surfaces $q = - m / n\,$. This analysis confirms  that {\em the trans-slow Alfv\'enic continuum modes are unstable at, or close to, the rational surfaces for all toroidal mode numbers $n$.}

For a very massive central object ($\Gamma \gg 1$), the growth rate in the limit $|n|\,, |m| \rightarrow \infty\,$ becomes
\BEQ \bar{\omega} \approx \pm \frac{i}{\sqrt{2\rho}} \,M \Gamma \,,\En{14}\EEQ
which is far in excess of the Alfv\'en frequency. 

Since these modes are {\em localized both radially} (because of the continuous spectrum) {\em and in the angle $\vartheta$ (and $\varphi$)} they are {\em perfectly suitable to produce the turbulence that is needed along the accretion flow at the inner edge of the accretion disk with respect to the central object to detach the flow from the magnetic field.}

\section{Conclusions}{\Sn{4}}

\noindent
(1) In the presence of poloidal rotation, the singular structure of the MHD continua transfers to the equilibrium, so that {\em linear waves and non-linear stationary equilibrium flows are no longer separate issues.}

\ver{2}\noindent
 (2) Complete spectra of waves and instabilities have been computed for tokamaks and accretion disks with transonic flows exploiting our new computational tools FINESSE (for equilibria) and PHOENIX (for stability). 

\ver{2}\noindent
 (3) We have found {\em a large class of instabilities of the continuous spectra of transonic axisymmetric equilibria for $M^2 > M_c^2$}. 

\ver{2}\noindent
 (4) These instabilities may cause strong MHD turbulence and associated anomalous dissipation, breaking the co-moving constraint of plasma and magnetic field and {\em facilitating both accretion and ejection of jets from accretion disks}.

\vskip 6 mm \noindent
{\bf Acknowledgements}

\vskip 3 mm
This work was performed as part of the research program of the Euratom-FOM Association Agreement, with support from the Netherlands Science Organization (NWO). The National Computing Facilities (NCF) is acknowledged for providing computer facilities.


\end{document}